\documentclass[reprint,onecolumn,
 amsmath,amssymb,
 aps,
prf,
]{revtex4-2}

\usepackage{graphicx}
\usepackage{dcolumn}
\usepackage{bm}
\usepackage{xcolor}
\usepackage{ulem}
\usepackage{float} 
\usepackage[frak = euler]{mathalpha}
\usepackage{hyperref}

\newcommand{\be}{\begin{equation}}
	\newcommand{\ee}{\end{equation}}
\newcommand{\bse}{\begin{subequations}}
	\newcommand{\ese}{\end{subequations}}
\newcommand{\bea}{\begin{eqnarray}}
	\newcommand{\eea}{\end{eqnarray}}
\newcommand{\ba}{\begin{array}}
	\newcommand{\ea}{\end{array}}
\newcommand{\mm}{mode-to-mode}
\newcommand{\ke}{kinetic energy}

\begin{document}


\title{Mathematical formulation of mode-to-mode energy transfers and energy fluxes in compressible turbulence}

\author{Dhananjay Singh}
 \email{sdhanjay@iitk.ac.in}
 
\author{Harshit Tiwari}%
 \email{tharshit@iitk.ac.in}
 \affiliation{ 
Department of Physics, Indian Institute of Technology Kanpur, Kanpur 208016, India
}

\author{Lekha Sharma}
 \email{lekhas@iitk.ac.in}
 \affiliation{ 
Department of Physics, Indian Institute of Technology Kanpur, Kanpur 208016, India
}

\author{Mahendra K. Verma}%
 \email{mkv@iitk.ac.in}
\affiliation{ 
Department of Physics, Indian Institute of Technology Kanpur, Kanpur 208016, India
}
\affiliation{ 
Kotak School of Sustainability, Indian Institute of Technology Kanpur, Kanpur 208016, India
}
\date{\today}

\begin{abstract}
Understanding compressible turbulence is critical for modeling atmospheric, astrophysical, and engineering flows. However, compressible turbulence poses a more significant challenge than incompressible turbulence. We present a novel mathematical framework to compute \textit{mode-to-mode energy transfer rates} and energy fluxes for compressible flows. The formalism captures detailed energy conservation within triads and allows decomposition of transfers into rotational, compressive, and mixed components, providing a clear picture of energy exchange among velocity and internal energy modes. We also establish analogies with incompressible hydrodynamic and magnetohydrodynamic flows, highlighting the framework’s universality in studying energy transfers. 

\end{abstract}

\maketitle


\section{INTRODUCTION}\label{sec:introduction}
Compressible turbulence~\cite{Lele:ARFM1994} is a fundamental phenomenon encountered in astrophysical applications~\cite{Carroll:book:Astrophysics}, such as solar convection~\cite{Spruit:ARAA1990}, astrophysical jets, and supernova explosions; in terrestrial atmospheres~\cite{Vallis:book}; and in engineering applications, including fusion plasmas~\cite{Fujisawa:PJA2021} and high-speed aerodynamics~\cite{Von:JSR2003}. In a fluid flow, the relative density fluctuation $(\delta \rho)/\rho $ varies as $(U/C_s)^2$, where $U$ is the velocity scale and $C_s$ is the sound speed. For example, a river flow is incompressible because $U \ll C_s$. However, fluid flows in combustion and astrophysical jets are compressible with $U \gtrapprox C_s$. Compressible flows involve complex interactions and multiscale energy transfers among the \textit{solenoidal} (or \textit{rotational}) and \textit{compressive} components of the velocity field, and internal energy. This paper constructs a mathematical framework to quantify triadic energy transfers and energy fluxes.

Compared to compressible turbulence, incompressible turbulence theory is well-established~\cite{Lesieur:book:Turbulence}. A key result in incompressible turbulence is Kolmogorov's spectrum,  $E(k) = K_\mathrm{Ko} \epsilon_\mathrm{Ko}^{2/3} k^{-5/3}$. This describes the kinetic energy distribution among wavenumbers ($k$) in the inertial range, where $\epsilon_\mathrm{Ko}$ is the constant energy flux and viscous dissipation rate, and $K_\mathrm{Ko}$ is Kolmogorov's constant~\cite{Lesieur:book:Turbulence,Kolmogorov:DANS1941Dissipation,Kolmogorov:DANS1941Structure}. Beyond the energy spectrum, the dynamics of scale interactions have also been explored in detail. 
\citet{Kraichnan:JFM1959} provided a framework of \textit{combined energy transfers} among the velocity Fourier modes in a wavenumber triad.  Later, 
\citet{Dar:PD2001} and \citet{Verma:PR2004} generalized \citet{Kraichnan:JFM1959}'s framework to \textit{mode-to-mode energy transfers}. Moreover, energy is predominantly exchanged between neighboring wavenumber shells—a property known as \textit{locality} of interactions—which underpins the scale-by-scale energy cascade in incompressible turbulence~\cite{Domaradzki:PF1990,Zhou:PF1993,Verma:Pramana2005S2S}. In contrast, the theoretical understanding of compressible turbulence is far less developed. The presence of an additional compressive velocity component and the coupling between the kinetic energy and internal energy via pressure-dilatation effects introduce significant complexity. As a result, the nature of interscale energy transfers, the role of compressibility, and the scaling laws remain active areas of research in compressible turbulence.

In the following discussion, we review past works on the energy spectra and transfers in compressible turbulence. Using numerical simulations, several researchers~\cite{Kida:JSC1990,Kida:JSC1992,Wang:PRL2013,Schmidt:PRE2019,Sakurai:PF2024} showed that the rotational velocity component exhibits Kolmogorov's $k^{-5/3}$ energy spectrum, whereas the compressive component exhibits $k^{-2}$ energy spectrum. The relative magnitude of compressive and rotational kinetic energies depends on the Mach number $\mathrm{Ma} = U/C_s$~\cite{Kida:JSC1990, Kida:JSC1992, Jagannathan:JFM2016, Sarkar:PF1992}. For example, using numerical simulations, \citet{Kida:JSC1990} showed that an increase of Ma leads to a relative increase in compressive energy. \citet{Kida:JSC1990,Kida:JSC1992}, \citet{Miura:POF1995}, and \citet{Praturi:POF2019} simulated compressible turbulence and computed the integrated energy transfers between kinetic and internal energies. \citet{Miura:POF1995} investigated the energy exchange between kinetic and internal energies and identified pressure-dilatation as the mechanism driving oscillatory energy transfers across scales. Their findings indicated that the oscillation period diminishes with increasing wavenumber, underscoring the role of acoustic modes in mediating scale-dependent energy exchange. \citet{Praturi:POF2019} noted that at high wavenumbers, compressive velocity fields carry more energy compared to their rotational counterparts, which is attributed to the lack of pressure action enforcing a divergence-free condition, thereby facilitating shock formation. \citet{Jagannathan:JFM2016} and \citet{John:PF2024_compress} investigated the scaling of global quantities with Reynolds and Mach numbers using numerical simulations and reported strong evidence supporting universality across different flow regimes. As an aside, for a nearly incompressible flow [$(\delta \rho)/\rho \rightarrow 0$], \citet{Zank:PF1991} showed that both velocity and density fields follow $k^{-5/3}$ spectrum (also see \cite{Verma:book:ET}). 

Several studies have examined interscale \ke~transfers in compressible turbulence using both coarse-graining and spectral techniques~\cite{Sagaut:book,Wang:PRL2013,Wang:JFM2018,Luo:arXiv2025,Schmidt:PRE2019,Praturi:POF2019,Graham:ApJ2010,Yang:JFM2021}. \citet{Aluie:PRL2011} argued that in the inertial range, the \ke~transfer is conservative and it is dominated by local interactions. \citet{Aluie:PhysD2013} derived scale-by-scale energy transfers in compressible turbulence. Similar relations have also been validated through simulations~\cite{Wang:JFM2018,Luo:arXiv2025,Schmidt:PRE2019,Graham:ApJ2010}. \citet{Wang:JFM2018} showed that the solenoidal energy flux is nearly independent of the turbulent Mach number $M_t$, whereas the compressible flux grows with $M_t$ because of the energy transfers from solenoidal to compressive modes. \citet{Luo:arXiv2025} decomposed subgrid-scale flux into solenoidal, compressible, and mixed-mode contributions. Though structure-function-based approaches have also been explored~\cite{Falkovich:JFM2010,Banerjee:JFM2014,Kritsuk:JFM2013}, a complete picture of kinetic energy exchange between the solenoidal and compressive modes remains incomplete.

\citet{Graham:ApJ2010} developed a spectral framework to study energy transfers in compressible magnetohydrodynamics turbulence. They derived the transfer terms and separated them into advective and compressive contributions. Applying this approach to simulations for solar surface convection, they identified vortex stretching against magnetic tension as the primary mechanism driving small-scale dynamo action. \citet{Grete:PP2017} and \citet{Schmidt:PRE2019} used the density-weighted velocity $\mathbf{w} = \sqrt{\rho} \, \mathbf{u}$ ($\rho$ is the density and \textbf{u} is the velocity) to analyze the kinetic energy transfers using implicit large-eddy simulations. They also decomposed the cascade into advective and compressive components and showed that the velocity spectra, scaling exponents, and energy fluxes depend strongly on the nature of the external forcing. Their results support a transition from Kolmogorov to Burgers scaling for velocity spectra depending on Ma and on forcing composition.

As described earlier, the energy spectra of the rotational and compressive velocity components consistently show $k^{-5/3}$ and $k^{-2}$ spectra, respectively. However, the multiscale energy transfers and fluxes for the compressible turbulence have not been computed in \textit{detail}. Building on earlier works of \citet{Graham:ApJ2010,Schmidt:PRE2019,Dar:PD2001}, and \citet{Verma:PR2004}, we derive formulas for the \textit{mode-to-mode transfers} and the energy fluxes for compressible turbulence, including the rotational and compressive velocity components and internal energy. The {mode-to-mode energy formalism} presented in this work generalizes the formalism used for incompressible turbulence \cite{Dar:PD2001,Verma:PR2004}.  In addition, several properties of the energy fluxes for compressible turbulence resemble those for magnetohydrodynamic turbulence~\cite{Verma:PR2004,Verma:book:ET}. We remark that our formulas have been derived from the first principles. 

This paper presents the theoretical framework of mode-to-mode energy transfer for compressible turbulence, including the derivation of exact expressions for spectral energy fluxes. It serves as the technical foundation for the companion Letter~\cite{Singh:PRL2025_submitted}, which focuses on numerical simulations for different turbulent Mach numbers ($M_t = 0.15, 0.30, 0.45$). There, we compute the energy fluxes using the formalism developed here, demonstrating nearly constant fluxes in both solenoidal and compressive components, along with energy conversion to internal energy via pressure dilatation.
 Note that we provide multiscale energy flux, which is more detailed than the cumulative energy transfers computed by 
\citet{Kida:JSC1992}, \citet{Miura:POF1995}, and \citet{Jagannathan:JFM2016}. These novel results reveal valuable insights into the dynamics of compressible turbulence and address the concerns raised in earlier works, e.g.,~\citet{Aluie:PRL2011}.

The structure of the paper is as follows: Section~\ref{sec:comp_eqns} presents the equations of compressible flows in spectral space. In Sec.~\ref{sec:mode_to_mode}, we derive the mode-to-mode energy transfers for compressible flows. We define the corresponding kinetic energy fluxes in Sec.~\ref{sec:Fluxes}. In Sec.~\ref{sec:comparison}, we compare our results with past ones. Finally, we conclude in Sec.~\ref{sec:summary}.

\section{Compressible equations in spectral space} \label{sec:comp_eqns}
The equations for a compressible flow in  nondimensional and tensorial form are~\cite{Kida:JSC1990}:
\begin{gather}
\frac{\partial \rho}{\partial t} + \frac{\partial}{\partial x_i}(\rho u_i)  = 0,\label{eq:continuity}  \\
\frac{\partial}{\partial t}(\rho u_i) + \frac{\partial}{\partial x_j}(\rho u_i u_j  + \delta_{ij} \sigma - \tau_{ij})  = \rho F_{i},\label{eq:momentum} \\
\frac{\partial E_T}{\partial t} + \frac{\partial}{\partial x_i} \left( u_i(E_T + \sigma) - \frac{1}{M_0^2 \mathrm{Pr} \mathrm{Re}_0 (\gamma -1)} \frac{\partial T}{\partial x_i} - u_j \tau_{ij} \right)  = \rho u_i F_i, \label{eq:energy}
\end{gather}
where $\rho, {\bf u},  \sigma, T, \mathbf{F}$ are the density, velocity, pressure, temperature, and  external force fields, respectively;  
\be
\tau_{ij} = \frac{1}{\mathrm{Re}_0} \left(\partial_j u_i + \partial_i u_j - \frac{2}{3} \partial_m u_m \delta_{ij} \right)
\ee
is the viscous stress tensor; and $E_T$  is the total energy density, which is a sum of  kinetic energy (KE) density 
\be
E_u = \frac{\rho u^2}{2}
\ee
and internal energy (IE) density 
\be
I = \frac{\sigma}{(\gamma - 1)},
\ee
with $\gamma = C_p/C_v$ as the ratio of specific heat capacities at constant pressure and volume. We assume the fluid to be an ideal gas with the equation of state
\be
\sigma = \frac{\rho T}{\gamma M_0^2}, 
\ee
where $M_0$ is the reference Mach number. Equations~(\ref{eq:continuity}-\ref{eq:energy}) have been nondimensionalized using reference density $\rho_0$, temperature $T_0$, velocity $u_0$, and length $l_0$.
The dimensionless parameters of the system are 
\bea
\mathrm{Reynolds~number}~~\mathrm{Re}_0 & = & \frac{\rho_0 u_0 l_0}{\mu} , \\
    \mathrm{Reynolds~number~(Taylor~microscale)}~~\mathrm{Re}_\lambda &=& \bigg(\frac{5}{3 \mu \epsilon } \bigg)^{1/2}  \rho_0 U^2, \\
    \mathrm{Mach~number}~~M_0 & = & \frac{u_0}{c} = \frac{u_0}{\sqrt{\gamma R^* T_0}}, \\
    \mathrm{Turbulent~Mach~number}~~M_t &=& \frac{U}{ c}, \\
    \mathrm{Prandtl~number}~~\mathrm{Pr} & = & \frac{\mu C_p}{K_c},
\eea
where $c$ is the sound speed; $R^*$ is the gas constant; $\mu$ is the dynamic viscosity; $\epsilon$ is the mean viscous dissipation rate; $U$ is the root-mean-square velocity; and $K_c$ is the thermal conductivity~\cite{Kida:JSC1990,Jagannathan:JFM2016}. When ${\bf F}=0$, an integration of Eq.~(\ref{eq:energy}) over space yields
\be
\frac{d}{dt} \int d{\bf r} E_T({\bf r},t) = 0.
\ee
This result implies that the total energy $E_T$ is conserved in the absence of an external force. Note, however, that the KE  is transferred to the IE. In this paper, we analyze the scale-by-scale transfers of KE and IE.

Fourier decomposition helps in the scale-by-scale analysis of energy contents and transfers in the flow~\cite{Verma:book:ET}. In the following discussion, we will derive the KE transfers among the velocity Fourier modes.  To make the KE quadratic, we rewrite it as 
as ${\bf u \cdot v }/2$ with
\be 
\mathbf{v} = \rho \mathbf{u}.
\ee 
Hence, the modal energy is~\cite{Graham:ApJ2010}
\be
E_u(\mathbf{k}) = \frac{1}{2} \Re\left[\mathbf{v}(\mathbf{k}) \cdot \mathbf{u}^*(\mathbf{k})\right]. \label{eq:modal_ke}
\ee
Fourier transform of Eqs.~(\ref{eq:continuity}, \ref{eq:momentum}) yields (see Appendix~\ref{app:ke_derv} for details)
\bea
\frac{d}{dt} \mathbf{v}(\mathbf{k}) & = &   -i \sum_{\mathbf{p}} \{\mathbf{k} \cdot \mathbf{u}(\mathbf{q})\} \mathbf{v}(\mathbf{p}) - i\mathbf{k}\sigma(\mathbf{k}) - \mathbf{d}(\mathbf{k}) + \mathbf{F'}(\mathbf{k}), \label{eq:vk} \\
\frac{d}{dt} \mathbf{u}(\mathbf{k}) & = &  -i \sum_{\mathbf{p}} \{\mathbf{p} \cdot \mathbf{u}(\mathbf{q})\} \mathbf{u}(\mathbf{p}) - \tilde{\pmb{\sigma}}(\mathbf{k}) - \tilde{\mathbf{d}}(\mathbf{k}) + \mathbf{F}(\mathbf{k}),\label{eq:uk}
\eea 
where $\mathbf{k} = \mathbf{p} + \mathbf{q}$, and
\be
\tilde{\pmb{\sigma}} = {\nabla}\sigma/{\rho};~~~\mathbf{d} = -\partial_j\tau_{ij},~~~\tilde{\mathbf{d}} = \mathbf{d}/\rho,~~~ \mathbf{F'} = \rho\mathbf{F}.
\ee
Using the above, we derive the following  dynamical equation for the modal KE, $E_u(\mathbf{k})$, 
\be
\frac{d}{dt} E_u(\mathbf{k}) = T_u(\mathbf{k}) - Q_{I,u}(\mathbf{k}) - D_{I,u}(\mathbf{k}) + \mathcal{F}_u({\bf k})
\label{eq:modal_ke_1},
\ee
where,
\bea 
T_u(\mathbf{k}) & = & \frac{1}{2} \sum_{\mathbf{p}} \mathrm{Im} \big[ \{\mathbf{k} \cdot \mathbf{u}(\mathbf{q})\} \{\mathbf{v}(\mathbf{p}) \cdot \mathbf{u}^*(\mathbf{k})\} 
+ \{\mathbf{p} \cdot \mathbf{u}(\mathbf{q})\} \{\mathbf{u}(\mathbf{p}) \cdot \mathbf{v}^*(\mathbf{k})\} \big], \label{eq:Tu} \\
Q_{I,u}(\mathbf{k}) & = & - \frac{1}{2} \mathrm{Im} \big[\sigma(\mathbf{k}) \{\mathbf{k} \cdot \mathbf{u}^*(\mathbf{k})\}\big] 
+ \frac{1}{2} \mathrm{Re} \big\{\tilde{\pmb{\sigma}}(\mathbf{k}) \cdot \mathbf{v}^*(\mathbf{k})\}, \label{eq:Qu} \\
D_{I,u}(\mathbf{k}) & = & \frac{1}{2}\mathrm{Re} \big[ {\mathbf{d}}(\mathbf{k})\cdot\mathbf{u}^*(\mathbf{k}) + \tilde{{\mathbf{d}}}(\mathbf{k})\cdot\mathbf{v}^*(\mathbf{k}) \big],\label{eq:Du} \\
\mathcal{F}_u(\mathbf{k}) & = & \frac{1}{2}\mathrm{Re}\big[ {\mathbf{F'}}(\mathbf{k})\cdot\mathbf{u}^*(\mathbf{k}) + {\mathbf{F}}(\mathbf{k})\cdot\mathbf{v}^*(\mathbf{k}) \big]. \label{eq:mathcal_F}
\eea
Interpretation of Eqs.~(\ref{eq:Tu}-\ref{eq:mathcal_F}) are as follows:
\begin{enumerate}
    \item $\sum_{\mathbf{p}} \mathrm{Im}\big[ \{\mathbf{k} \cdot \mathbf{u}(\mathbf{q})\} \{\mathbf{v}(\mathbf{p}) \cdot \mathbf{u}^*(\mathbf{k})\}\big]$: KE transfer from ${\bf v(p)}$ to ${\bf u(k)}$.
    \item $\sum_{\mathbf{p}} \mathrm{Im}\big[\{\mathbf{p} \cdot \mathbf{u}(\mathbf{q})\} \{\mathbf{u}(\mathbf{p}) \cdot \mathbf{v}^*(\mathbf{k})\} \big]$: KE transfer from ${\bf u(p)}$ to ${\bf v(k)}$.
    \item $Q_{I,u}(\mathbf{k})$: At wavenumber \textbf{k}, transfer of KE to IE via pressure-dilatation or the work done by pressure~\cite{Kundu:book}.
    \item $D_{I,u}(\mathbf{k})$: Viscous dissipation of KE at wavenumber \textbf{k}. This lost KE reaches IE.
    \item $\mathcal{F}_u(\mathbf{k})$: At wavenumber \textbf{k},  KE injection  rate by the external force \textbf{F}.
\end{enumerate}
The unit interactions of items 1 and 2 involve wavenumbers of a triad, but those in items 3, 4, and 5 include wavenumbers beyond those in the triad because $\sigma({\bf k}), \mathbf{d}({\bf k}), \tilde{\pmb{\sigma}}(\mathbf{k}), \tilde{{\mathbf{d}}}(\mathbf{k})$, and ${\mathbf{F}'}(\mathbf{k})$ involve convolutions.

We focus on a single triad without ${\bf F}$ to obtain detailed scale-by-scale energy transfers. Following \citet{Kraichnan:JFM1959,Dar:PD2001}, and \citet{Verma:PR2004}, we consider a triad $({\bf k',p,q})$ with ${\bf k'+p+q}=0$~\cite{Dar:PD2001,Verma:book:ET}. Note that ${\bf k'= -k}$. For this triad, 
\be
\frac{d}{dt} E_u(\mathbf{k'}) = S^{uu}(\mathbf{k'|p,q}) - Q_{I,u}(\mathbf{k'}) 
 - D_{I,u}(\mathbf{k'})  
\label{eq:modal_ke_2},
\ee
where
\bea
S^{uu}(\mathbf{k}'|\mathbf{p}, \mathbf{q}) = -\frac{1}{2} \mathrm{Im}\big[\{\mathbf{k}' \cdot \mathbf{u}(\mathbf{q})\} \{\mathbf{v}(\mathbf{p}) \cdot \mathbf{u}(\mathbf{k}')\} 
- \{\mathbf{p} \cdot \mathbf{u}(\mathbf{q})\} \{\mathbf{u}(\mathbf{p}) \cdot \mathbf{v}(\mathbf{k}')\}\big] + \mathbf{p} \leftrightarrow \mathbf{q}
\label{eq:combined_ET}
\eea
is the \textit{combined KE transfer} to wavenumber $\mathbf{k'}$ from wavenumbers \textbf{p} and \textbf{q}. We can show that (see Appendix~\ref{app:ke_derv} for a proof)
\be
S^{uu}(\mathbf{k}'|\mathbf{p}, \mathbf{q}) + S^{uu}(\mathbf{p}|\mathbf{k}', \mathbf{q}) + S^{uu}(\mathbf{q}|\mathbf{p}, \mathbf{k}') = 0.
\ee
Therefore, 
\bea
\frac{d}{dt} [E_u({\bf k'}) +E_u({\bf p})+E_u({\bf q})] & = & -[Q_{I,u}({\bf k'}) + Q_{I,u}({\bf p}) + Q_{I,u}({\bf q})  +D_{I,u}({\bf k'}) + D_{I,u}({\bf p}) + D_{I,u}({\bf q})] \ne 0 .
\label{eq:triadic_int}
\eea
The right-hand-side of Eq.~(\ref{eq:triadic_int}) does not contain  $S^{uu}(\mathbf{k}'|\mathbf{p}, \mathbf{q})$, implying that the net triadic KE is conserved for the triadic interactions.  This is \textit{detailed conservation law} for compressible hydrodynamics, which a generalization of a similar law for incompressible hydrodynamics~\cite{Kraichnan:JFM1959}. Note, however, that the net KE of the triad is not conserved as $Q_{I,u}({\bf k'})$ and $D_{I,u}({\bf k'})$  transfer  KE to IE. 

In the next section, we will formulate \textit{mode-to-mode} energy transfers in compressible turbulence.

\section{Mode-to-mode energy transfers in compressible turbulence} \label{sec:mode_to_mode}

As shown in Sec.~\ref{sec:comp_eqns}, for a triad (${\bf k',p,q}$), $S^{uu}({\bf k'|p,q})$ of Eq.~(\ref{eq:combined_ET}) provides the net KE transfer from wavenumbers \textbf{p} and \textbf{q} to wavenumber $\mathbf{k'}$. In this section, we derive individual KE transfer from wavenumber \textbf{p} to wavenumber $\mathbf{k'}$ and from wavenumber \textbf{q} to wavenumber $\mathbf{k'}$. In this section,  we prove that for a triad $({\bf a,b,c})$ with ${\bf a+b+c} =0$, the \textit{mode-to-mode kinetic energy transfer}  from wavenumber \textbf{b} to wavenumber \textbf{a} with the mediation of wavenumber \textbf{c} is given by
\be
S^{uu}(\mathbf{a}|\mathbf{b}|\mathbf{c}) = -\frac{1}{2} \mathrm{Im}\big[\{\mathbf{a} \cdot \mathbf{u}(\mathbf{c})\} \{\mathbf{v}(\mathbf{b}) \cdot \mathbf{u}(\mathbf{a})\} 
- \{\mathbf{b} \cdot \mathbf{u}(\mathbf{c})\} \{\mathbf{u}(\mathbf{b}) \cdot \mathbf{v}(\mathbf{a})\}\big].
\label{Eq:M2M_uu}
\ee
 This formula is a generalization of a similar framework for incompressible turbulence derived by \citet{Dar:PD2001} and \citet{Verma:PR2004}.

\begin{figure}[h]
\includegraphics[width=0.55\textwidth]{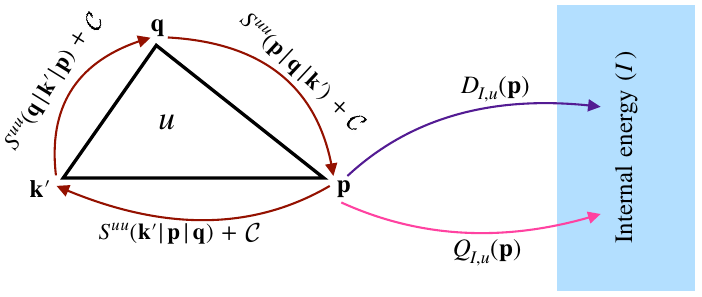}
\caption{Schematic  illustration of mode-to-mode energy transfers  $S^{uu}({\bf k'|p|q})$ in a triad $({\bf k',p,q})$. Here,  $Q_{I,u}({\bf p})$ and $D_{I,u}({\bf p})$ represent pressure dilatation and viscous dissipation, respectively. $\mathcal{C}$ is the arbitrary circulatory transfer that can be added to $S^{uu}({\bf k'|p|q})$.}
\label{fig:mode_u}
\end{figure}

We prove the above statement [Eq.~(\ref{Eq:M2M_uu})] using a similar approach to that used for incompressible turbulence. By definition, the mode-to-mode KE transfer formula must satisfy the following properties:
\begin{enumerate}
    \item The combined energy transfer is a sum of individual mode-to-mode energy transfers, i.e.,  
    \bea
    S^{uu}({\bf a|b|c}) + S^{uu}({\bf a|c|b}) & = & S^{uu}({\bf a|b,c}).\label{Eq:M2M_P1} 
    \eea
    The above property is a logical consequence of its definition [Eq.~(\ref{eq:combined_ET})]. Similarly, we derive two more equations for $S^{uu}({\bf b|a,c})$ and $S^{uu}({\bf c|a,b})$.
    
    \item The mode-to-mode transfer from wavenumber \textbf{a} to \textbf{b} is equal and opposite to that from \textbf{b} to \textbf{a}, i.e.,
    \be
    S^{uu}({\bf a|b|c}) = - S^{uu}({\bf b|a|c}).\label{Eq:M2M_P2}
    \ee 
    We prove the above statement as follows:
    \bea
    S^{uu}({\bf a|b|c}) + S^{uu}({\bf b|a|c})
    & = & -\frac{1}{2} \mathrm{Im}\big[\{\mathbf{a} \cdot \mathbf{u}(\mathbf{c})\} \{\mathbf{v}(\mathbf{b}) \cdot \mathbf{u}(\mathbf{a})\} \nonumber \\
&& - \{\mathbf{b} \cdot \mathbf{u}(\mathbf{c})\} \{\mathbf{u}(\mathbf{b}) \cdot \mathbf{v}(\mathbf{a})\} \nonumber \\
&& + \mathbf{b} \cdot \mathbf{u}(\mathbf{c})\} \{\mathbf{v}(\mathbf{a}) \cdot \mathbf{u}(\mathbf{b})\} \nonumber \\
&& - \{\mathbf{a} \cdot \mathbf{u}(\mathbf{c})\} \{\mathbf{u}(\mathbf{a}) \cdot \mathbf{v}(\mathbf{b})\}
\big] = 0.
    \eea
    There are two additional equations for $S^{uu}({\bf b|c|a})$ and $S^{uu}({\bf c|a|b})$.
\end{enumerate}
Thus, we have 6 equations for six unknowns $S^{uu}({\bf a|b|c})$, $S^{uu}({\bf a|c|b})$, $S^{uu}({\bf b|a|c})$, $S^{uu}({\bf b|c|a})$, $S^{uu}({\bf c|a|b})$, $S^{uu}({\bf c|b|a})$. It is easy to verify that $S^{uu}({\bf a|b|c})$ of the form given in Eq.~(\ref{Eq:M2M_uu}) satisfies Eqs.~(\ref{Eq:M2M_P1}, \ref{Eq:M2M_P2}). Therefore, $S^{uu}({\bf a|b|c})$ is indeed the mode-to-mode transfer from wavenumber \textbf{b} to wavenumber \textbf{a} with the mediation of wavenumber \textbf{c}. Unfortunately, the above solution is not unique. Following \citet{Dar:PD2001}, we can add a circulatory transfer $\mathcal{C}$ to the formula (see Fig.~\ref{fig:mode_u}). Fortunately, these circulatory transfers do not affect the energy flux, and hence, we can set $\mathcal{C} = 0$, just like a gauge field in electromagnetism~\cite{Dar:PD2001}.   Figure~\ref{fig:mode_u} illustrates the mode-to-mode energy transfers for a triad $({\bf k', p, q}) $. It also exhibits the loss of modal KE, $E_u({\bf p})$, to IE via pressure dilatation $Q_{I,u}({\bf p})$ and viscous dissipation $D_{I,u}({\bf p})$.

\begin{figure}[h]
\includegraphics[width=0.32\textwidth]{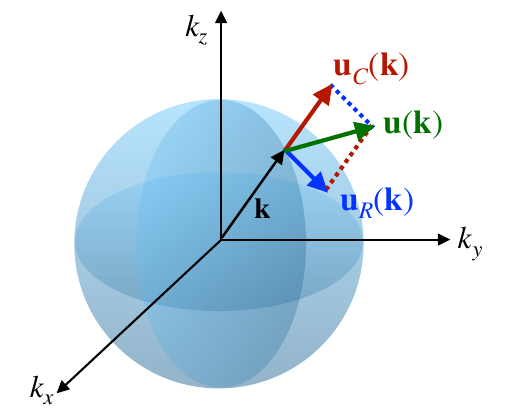}
\caption{Decomposition of a velocity mode $\mathbf{u(k)}$ into its rotational component $\mathbf{u}_R({\bf k})$ and compressive component $\mathbf{u}_C({\bf k})$.} \label{fig:u_comp}
\end{figure}

We gain further insights into the system by dividing the velocity field into its rotational (or solenoidal) component $\mathbf{u}_R$ and compressive component $\mathbf{u}_C$~\cite{Kida:JSC1990}. In Fourier space, 
\be
\mathbf{u}(\mathbf{k}) = \mathbf{u}_R(\mathbf{k}) + \mathbf{u}_C(\mathbf{k}).
\ee
Note that $\mathbf{u}_C(\mathbf{k})$ is along \textbf{k} and $\mathbf{u}_R(\mathbf{k})$ is perpendicular to \textbf{k}, as illustrated in Fig.~\ref{fig:u_comp}. The modal kinetic energies for these components are
\be
E_{\alpha}(\mathbf{k}) = \frac{1}{2} \mathrm{Re}\big[\mathbf{v}_{\alpha}(\mathbf{k}) \cdot \mathbf{u}_{\alpha}^*(\mathbf{k})\big],
\ee
where $\alpha =R,C$, whereas the total kinetic energy is
\be
E_u({\bf k}) = E_R({\bf k})+E_C({\bf k}) =
\frac{1}{2} \mathrm{Re}\big[\mathbf{v}_{R}(\mathbf{k}) \cdot \mathbf{u}_{R}^*(\mathbf{k})\big]
+ \frac{1}{2} \mathrm{Re}\big[\mathbf{v}_{C}(\mathbf{k}) \cdot \mathbf{u}_{C}^*(\mathbf{k})\big].
\ee
The  dynamical equation for $E_\alpha(\mathbf{k}')$ is
\bea
\frac{d}{dt} E_\alpha(\mathbf{k}')
 & = & T_{\alpha \alpha} + T_{\alpha \beta}- Q_{I, \alpha}(\mathbf{k}') \nonumber - D_{I, \alpha}(\mathbf{k}')
 + \mathcal{F}_{\alpha}(\mathbf{k}')\\
 & = &  \sum_{\bf p} S^{\alpha \alpha}({\bf k'|p|q}) + \sum_{\bf p} S^{\alpha \beta}({\bf k'|p|q}) - Q_{I,\alpha}(\mathbf{k}')
 - D_{I,\alpha}(\mathbf{k}')
 + \mathcal{F}_{\alpha}(\mathbf{k}'),
 \label{eq:dt_E_alpha_k}
\eea
where 
\bea
S^{\alpha \alpha}(\mathbf{k}'|\mathbf{p}|\mathbf{q}) & = & -\frac{1}{2} \mathrm{Im}\big[\{\mathbf{k}' \cdot \mathbf{u}(\mathbf{q})\} \{\mathbf{v}_\alpha (\mathbf{p}) \cdot \mathbf{u}_\alpha (\mathbf{k}')\} 
- \{\mathbf{p} \cdot \mathbf{u}(\mathbf{q})\} \{\mathbf{u}_\alpha (\mathbf{p}) \cdot \mathbf{v}_\alpha (\mathbf{k}')\}\big],\\
S^{\alpha \beta}(\mathbf{k}'|\mathbf{p}|\mathbf{q}) & = & -\frac{1}{2} \mathrm{Im}\big[\{\mathbf{k}' \cdot \mathbf{u}(\mathbf{q})\} \{\mathbf{v}_{\beta}(\mathbf{p}) \cdot \mathbf{u}_\alpha (\mathbf{k}')\} 
- \{\mathbf{p} \cdot \mathbf{u}(\mathbf{q})\}\{\mathbf{u}_{\beta}(\mathbf{p}) \cdot \mathbf{v}_\alpha (\mathbf{k}')\}\big], \\
Q_{I,R}(\mathbf{k}') & = & \frac{1}{2} \mathrm{Re}\big[\tilde{\pmb{\sigma}}(\mathbf{k}') \cdot \mathbf{v}_R^*(\mathbf{k}')\big], \\
Q_{I,C}(\mathbf{k}') & = & \frac{1}{2} \mathrm{Re}\big[\tilde{\pmb{\sigma}}(\mathbf{k}') \cdot \mathbf{v}_C^*(\mathbf{k}')\big] 
- \frac{1}{2} \mathrm{Im}\big[\sigma(\mathbf{k}') \{\mathbf{k}' \cdot \mathbf{u}_C^*(\mathbf{k}') \}\big], \\
D_{I,\alpha}(\mathbf{k}') & = & \frac{1}{2} \mathrm{Re}\big[\mathbf{d}_{\alpha}(\mathbf{k}') \cdot \mathbf{u}_{\alpha}^*(\mathbf{k}') + \tilde{\mathbf{d}}_{\alpha}(\mathbf{k}') \cdot \mathbf{v}_{\alpha}^*(\mathbf{k}') \big], \\
\mathcal{F}_{\alpha}(\mathbf{k}') & = & \frac{1}{2} \mathrm{Re}\big[\mathbf{F'}_{\alpha}(\mathbf{k}') \cdot \mathbf{u}_{\alpha}^*(\mathbf{k}') + \mathbf{F}_{\alpha}(\mathbf{k}') \cdot \mathbf{v}_{\alpha}^*(\mathbf{k}') \big]. 
\eea
Note that
\bea
S^{\alpha \alpha}({\bf k'|p|q}) + S^{\alpha \alpha}({\bf p|k'|q}) & = & 0,
\label{eq:Suu_alpha_zero} \\
S^{\alpha \beta}({\bf k'|p|q}) + S^{\beta \alpha}({\bf p|k'|q}) & = & 0.
\label{eq:Suu_alpha__beta_zero}
\eea

\begin{figure}[h]
\includegraphics[width=0.5\textwidth]{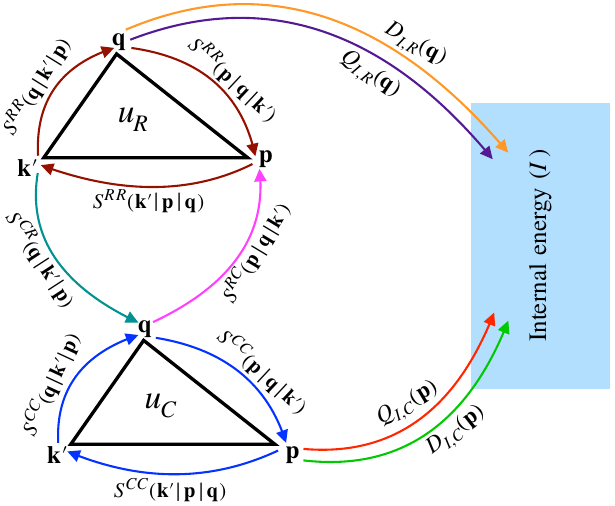}
\caption{Mode-to-mode energy transfer terms between decomposed kinetic energies with rotational and compressive velocity components, ${\bf u}_R$ and ${\bf u}_C$, respectively: $S^{\alpha \alpha}({\bf k'|p|q})$ represents energy transfer rate from mode ${\bf u}_\alpha ({\bf p})$ to mode ${\bf u}_\alpha ({\bf k'})$ with the mediation of full ${\bf u} ({\bf q})$, while $S^{\alpha \beta}({\bf k'|p|q})$ represents energy transfer rate from mode ${\bf u}_\beta ({\bf p})$ to mode ${\bf u}_\alpha ({\bf k'})$ with the mediation of full ${\bf u} ({\bf q})$. $Q_{I,\alpha}({\bf p})$ and $D_{I,\alpha}({\bf p})$ denote pressure dilatation and viscous dissipation of KE at wavenumber ${\bf p}$ to internal energy ($I$). Here, $\alpha,\beta = R,C$ where $R$ and $C$ represent the rotational and compressive kinetic energy modes.}\label{fig:mode}
\end{figure}

Interestingly, $S^{\alpha \alpha}({\bf k'|p|q})$ and its variants satisfy Eqs.~(\ref{Eq:M2M_P1},\ref{Eq:M2M_P2}). Therefore, following the same arguments as before, we can deduce that for the component $\alpha$ ($R$ or $C$), $S^{\alpha \alpha}({\bf k'|p|q})$ represents the energy transfer rate from wavenumber \textbf{p} to wavenumber \textbf{k}' with the mediation of wavenumber \textbf{q}. Here, the giver and receiver modes belong to $\mathbf{u}_\alpha$ field, but the mediator is the full ${\bf u}$ field.  We illustrate these energy transfers in Fig.~\ref{fig:mode}. Using Eq.~(\ref{eq:Suu_alpha_zero}), we derive that
\begin{align}
S^{\alpha \alpha}({\bf k'|p|q}) + S^{\alpha \alpha}({\bf k'|q|p}) + S^{\alpha \alpha}({\bf p|q|k'}) & \\ \nonumber 
+ S^{\alpha \alpha}({\bf p|k'|q}) +  S^{\alpha \alpha}({\bf q|k'|p}) + S^{\alpha \alpha}({\bf q|p|k'}) &= 0,
\end{align}
which implies that $E_\alpha({\bf k'}) + E_\alpha({\bf p})+E_\alpha({\bf q})$ is \textit{partially} conserved for the energy transfers along this channel.

In addition, $S^{\alpha \beta}({\bf k'|p|q})$ and its variants satisfy Eqs.~(\ref{Eq:M2M_P1},\ref{eq:Suu_alpha__beta_zero}). Therefore, using similar reasoning, we interpret $S^{\alpha \beta}({\bf k'|p|q})$ as the energy transfer rate from  ${\bf u}_\beta({\bf p})$ to  ${\bf u}_\alpha({\bf k'})$ with the mediation of ${\bf u(q)}$.  Here, the giver and receiver modes belong to $\mathbf{u}_\beta$ and $\mathbf{u}_\alpha$ fields, respectively, but the mediator is the full ${\bf u}$ field. Here, $\alpha \ne \beta$. These cross transfers are illustrated in Fig.~\ref{fig:mode}. Note that an application of Eq.~(\ref{eq:Suu_alpha__beta_zero}) yields
\begin{align}
S^{\alpha \beta}({\bf k'|p|q}) + S^{\alpha \beta}({\bf k'|q|p}) + S^{\alpha \beta}({\bf p|q|k'}) & \\ \nonumber 
+ S^{\alpha \beta}({\bf p|k'|q}) +  S^{\alpha \beta}({\bf q|k'|p}) + S^{\alpha \beta}({\bf q|p|k'}) & \\ \nonumber
+ S^{\beta \alpha}({\bf k'|p|q}) + S^{\beta \alpha}({\bf k'|q|p}) + S^{\beta \alpha}({\bf p|q|k'}) & \\ \nonumber 
+ S^{\beta \alpha}({\bf p|k'|q}) +  S^{\beta \alpha}({\bf q|k'|p}) + S^{\beta \alpha}({\bf q|p|k'}) &= 0,
\end{align}
which implies a conservation of KE by these interactions. The above transfers redistribute the KE among the triadic modes without any gain or loss. We can also write the mode-to-mode transfer terms in terms of $({\bf k,p,q})$ with ${\bf k = p + q}$ as
\bea
    S^{\alpha \alpha}(\mathbf{k}|\mathbf{p}|\mathbf{q}) & = & \frac{1}{2} \mathrm{Im}\big[\{\mathbf{k} \cdot \mathbf{u}(\mathbf{q})\} \{\mathbf{v}_{\alpha}(\mathbf{p}) \cdot \mathbf{u}^*_\alpha (\mathbf{k})\} 
+ \{\mathbf{p} \cdot \mathbf{u}(\mathbf{q})\}\{\mathbf{u}_{\alpha}(\mathbf{p}) \cdot \mathbf{v}^*_\alpha (\mathbf{k})\}\big], \\
    S^{\alpha \beta}(\mathbf{k}|\mathbf{p}|\mathbf{q}) & = &  \frac{1}{2} \mathrm{Im}\big[\{\mathbf{k} \cdot \mathbf{u}(\mathbf{q})\} \{\mathbf{v}_{\beta}(\mathbf{p}) \cdot \mathbf{u}^*_\alpha (\mathbf{k})\} 
+ \{\mathbf{p} \cdot \mathbf{u}(\mathbf{q})\}\{\mathbf{u}_{\beta}(\mathbf{p}) \cdot \mathbf{v}^*_\alpha (\mathbf{k})\}\big].
\eea

At this point, it is important to compare the above formulas with those for incompressible flows, where $\mathbf{u}_C = 0$ and $\rho = $ const. Since $\mathbf{u}_C=0$, we easily deduce that
\be
S^{C C}({\bf k|p|q}) = 0;~~~Q_{I,C}({\bf k}) = 0, 
\ee
and
\bea
S^{R R}(\bf k|p|q) & = & \frac{\rho}{2}  \mathrm{Im}\big[\{\mathbf{k} \cdot \mathbf{u}_R(\mathbf{q})\} \{\mathbf{u}_R(\mathbf{p}) \cdot \mathbf{u}_R^*(\mathbf{k})\} + \{(\mathbf{k-q}) \cdot \mathbf{u}_R(\mathbf{q})\} \{\mathbf{u}_R(\mathbf{p}) \cdot \mathbf{u}_R^*(\mathbf{k})\} \big], \nonumber \\
& = & \rho \mathrm{Im}\big[\{\mathbf{k} \cdot \mathbf{u}_R(\mathbf{q})\} \{\mathbf{u}_R(\mathbf{p}) \cdot \mathbf{u}_R^*(\mathbf{k})\}, 
\eea
which is same as that derived by \citet{Dar:PD2001}. Similarly, we show that
\bea
Q_{I,R}({\bf k})  & = &  \frac{1}{2}\mathrm{Re}\big[i\sigma(\mathbf{k})\{\mathbf{k}\cdot \mathbf{u}^*(\mathbf{k})\}\big] = 0
\eea
for an incompressible flow. Note that the internal energy, $I$, is not considered in an incompressible flow.  It is implicitly assumed that the viscous dissipation increases the internal energy, but this conversion is not accounted for. Thus,  the energy transfers in compressible turbulence reduce to those in incompressible turbulence where ${\bf u}_C=0$ and $\rho$ is constant.

In this section, we discussed the energy transfer formalism with the kinetic energy density as ${\bf (u \cdot v)}/2$,  where ${\bf v = \rho u}$ is the new variable. There is an equivalent framework where the kinetic energy density is ${\bf w \cdot w}/2  $ with ${\bf w } = \sqrt{\rho} {\bf u}$~\cite{Kida:JSC1990, Kida:JSC1992, Schmidt:PRE2019, Grete:PP2017}.  This alternate framework is described in  Appendix \ref{app:mode_w}.  In the following section, Sec.~\ref{sec:Fluxes}, we describe the energy fluxes in compressible turbulence.

\section{Energy fluxes in compressible turbulence}\label{sec:Fluxes}

The mode-to-mode energy transfers described in earlier sections provide valuable insights into the turbulence dynamics. However, these transfers exhibit substantial fluctuations, obscuring underlying patterns. Therefore, researchers employ energy flux,  shell-to-shell transfers, and ring-to-ring transfers, which involve sums of many unit energy transfers [$S^{\alpha \alpha}(\bf k|p|q)$ and $S^{\alpha \beta}(\bf k|p|q)$]~\cite{Verma:book:ET}. In this section, we derive the energy fluxes in compressible turbulence.

\begin{figure}[h]
\includegraphics[width=0.55\textwidth]{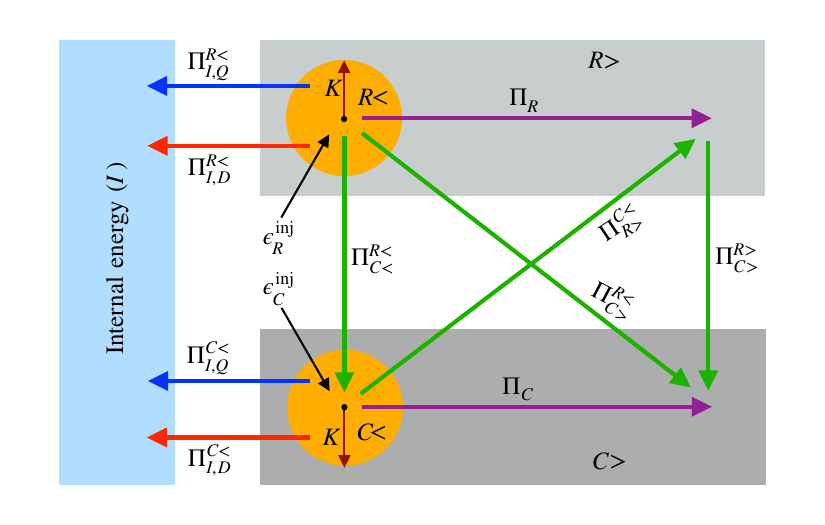}
\caption{Various fluxes in compressible turbulence: Rotational and compressive energy fluxes ($\Pi_{R}$, $\Pi_{C}$), cross fluxes ($\Pi^{R< }_{C<}, \Pi^{R< }_{C>}, \Pi^{R> }_{C>}, \Pi^{C< }_{R>}$), pressure dilatation ($\Pi^{R}_{I,Q}, \Pi^{C}_{I,Q}$), viscous dissipation ($\Pi^{R}_{I,D}, \Pi^{C}_{I,D}$), and energy injection rates ($\epsilon^\mathrm{inj}_R, \epsilon^\mathrm{inj}_C$). }
\label{fig:flux_comp}
\end{figure}

Using the \mm\ energy transfers, we  define energy transfers from ${\bf u}_\beta$ in the wavenumber region $Y$ to ${\bf u}_\alpha$ in the wavenumber region $X$ as~\cite{Verma:PR2004,Verma:book:ET} 
\be
\mathcal{T}^{\beta,Y}_{\alpha,X} = \sum_{{\bf k} \in X} \sum_{ {\bf p} \in Y} S^{\alpha \beta}({\bf k|p|q}) .
\label{eq:ET_XY}
\ee
In $\mathcal{T}^{\beta,Y}_{\alpha,X}$, the superscript refers to the giver field, whereas the subscript refers to the receiver field. Using this broad framework, we define the energy flux for the rotational and compressive components as follows. The energy flux $\Pi_{R}(K)$ [$\Pi_{C}(K)$] is the net energy transfer from the ${\bf u}_R$ (${\bf u}_C$) modes inside the wavenumber sphere  to the ${\bf u}_R$ (${\bf u}_C$) modes outside the sphere of radius $K$, i.e.,
\bea
\Pi_{R}(K) = \Pi^{R <}_{R >}(K) & = &  \sum_{k>K} \sum_{p \le K} S^{R R}({\bf k|p|q}), \\
\Pi_{C}(K) = \Pi^{C <}_{C >}(K) & = &  \sum_{k>K} \sum_{p \le K} S^{C C}({\bf k|p|q}).
\eea
Here,  $<$ denotes modes within the wavenumber sphere, and $>$ denotes modes outside it. In addition to the above, we define fluxes for the  cross transfers  between ${\bf u}_R$ and ${\bf u}_C$:
\bea
\Pi^{R <}_{C <}(K) & = &  \sum_{k \le K} \sum_{p \le K} S^{C R}({\bf k|p|q}) 
\label{eq:rc_ll},
\\
\Pi^{R >}_{C >}(K) & = &  \sum_{k>K} \sum_{p> K} S^{C R}({\bf k|p|q}), \\
\Pi^{R <}_{C >}(K) & = & \sum_{k>K} \sum_{p\le K} S^{C R}({\bf k|p|q}), \\
\Pi^{C <}_{R >}(K) & = &  \sum_{k>K} \sum_{p \le K} S^{R C}({\bf k|p|q}).
\label{eq:cr_lg}
\eea
For example, $\Pi^{R <}_{C >}(K)$ is the energy transfers from $\mathbf{u}_R$ modes inside the sphere of radius $K$ to $\mathbf{u}_C$ modes outside the sphere, whereas  $\Pi^{R <}_{C <}(K)$ is the energy transfers from $\mathbf{u}_R$ modes to $\mathbf{u}_C$ modes inside the sphere.  These fluxes are illustrated in Fig.~\ref{fig:flux_comp}. We  also define the energy fluxes using the nonlinear energy transfers $T_{\alpha \alpha}({\bf k})$ and $T_{\alpha \beta}({\bf k})$ as follows~\cite{Lesieur:book:Turbulence,Verma:book:ET,Verma:JPA2022}:
\begin{align}
\Pi_{\alpha}(K) & = -\sum_{k \leq K}
T_{\alpha \alpha}({\bf k}),
\label{eq:flux_self_T} \\
\Pi^{\alpha <}_{\beta}(K) & = \Pi^{\alpha <}_{\beta<}(K) + \Pi^{\alpha <}_{\beta>}(K) = -\sum_{k \leq K}
T_{\alpha \beta}({\bf k}) .
\label{eq:flux_cross_T}
\end{align}
Equation~(\ref{eq:flux_cross_T}) represents the net energy transfers from ${\bf u}_\alpha$ modes inside the sphere to all the ${\bf u}_\beta$ modes.  Note that $\Pi_{\alpha}(K=\infty) = 0$   because 
\be
\Pi_{\alpha}(K=\infty) = \sum_{0<k'<\infty} \sum_{0<p<\infty} S^{\alpha \alpha}({\bf k'|p|q}) = \frac{1}{2} \sum_{0<k'<\infty} \sum_{0<p<\infty}   [S^{\alpha \alpha}({\bf k'|p|q})+S^{\alpha \alpha}({\bf p|k'|q})] = 0.
\ee
We use Eqs.~(\ref{eq:flux_self_T}, \ref{eq:flux_cross_T}) to calculate energy fluxes efficiently.  A fraction of kinetic energy is transferred to the internal energy via pressure dilatation $Q$ and viscous dissipation $D$, for which we define the energy fluxes $\Pi^{{\alpha} <}_{I,Q}(K)$ and $ \Pi^{{\alpha} <}_{I,D}(K)$ as follows:
\bea
\Pi^{{\alpha} <}_{I,Q}(K) & = &   \sum_{k \le K} Q_{I,\alpha}(\mathbf{k}), \\
\Pi^{{\alpha} <}_{I,D}(K) & = &   \sum_{k \le K} D_{I,\alpha}(\mathbf{k}).
\eea

In our numerical simulations, discussed in the companion Letter~\cite{Singh:PRL2025_submitted}, we show that $Q_{I,{\alpha}}(\mathbf{k})$ transfers are dominant at large scales, but viscous dissipation $D_{I,{\alpha}}(\mathbf{k})$ occur at all scales. We illustrate these transfers in Fig.~\ref{fig:flux_comp}.  Using the aforementioned formulas, we write the following equations:
   \bea
\frac{d}{dt} \sum_{k \leq K} E_R({\bf k}) & = & -\Pi_{R}(K) - \Pi^{R<}_{C}(K)-  \Pi^{R <}_{I,Q}(K)
-\Pi^{R <}_{I,D}(K) + \epsilon^\mathrm{inj}_R, \\
\frac{d}{dt} \sum_{k \leq K} E_C({\bf k}) & = & -\Pi_{C}(K) - \Pi^{C<}_{R}(K) -  \Pi^{C <}_{I,Q}(K)
-\Pi^{C <}_{I,D}(K) + \epsilon^\mathrm{inj}_C,
\eea
with
\bea 
\sum_{k\le k_f} \mathcal{F}_R(k)& = &\epsilon^\mathrm{inj}_R, \\
\sum_{k\le k_f} \mathcal{F}_C(k)& = &\epsilon^\mathrm{inj}_C,
\eea 
where $\epsilon^\mathrm{inj}_R$ and  $\epsilon^\mathrm{inj}_C$ are the energy injection rates to the rotational and compressive components, respectively; and $k_f$ is the forcing wavenumber band.  For a steady state, $d/dt \sum_{k \leq K} E_\alpha({\bf k}) = 0$, we obtain the following exact relations for $K > k_f$:
\bea
\Pi_{R}(K) + \Pi^{R<}_{C}(K) + \Pi^{R<}_{I,Q}(K) + \Pi^{R<}_{I,D}(K) & = & \epsilon^\mathrm{inj}_R = I_1 = \mathrm{const.};
\label{eq:I1} \\
\Pi_{C}(K) + \Pi^{C<}_{R}(K) + \Pi^{C<}_{I,Q}(K) + \Pi^{C<}_{I,D}(K) & = & \epsilon^\mathrm{inj}_C = I_2= \mathrm{const.}
\label{eq:I2}
\eea
These exact relations, which are similar to those for MHD turbulence and similar systems~\cite{Verma:JPA2022,Verma:book:ET, Verma:Fluid2021},  hold for  $K$'s in the inertial-dissipation range (beyond the forcing wavenumbers).

Equation~(\ref{eq:ET_XY}) and the energy flux formulas are similar to those used for incompressible flows, such as magnetohydrodynamic turbulence~\cite{Dar:PD2001,Verma:book:ET}.
This similarity is because of the accurate identification of the giver and receiver modes that capture the common features of energy transfers irrespective of details. The formulas for the mode-to-mode energy transfers for the compressible and incompressible turbulence are different in form, but they are very similar in spirit. Hence, the above energy transfer framework provides valuable universal tools for turbulence analysis. 


\section{Comparison with Past works}\label{sec:comparison}

Finally, we compare our results with those of past works. In 1990, \citet{Kida:JSC1990} simulated compressible turbulence on $64^3$ grid and computed energy transfers among the rotational and compressive velocity components and the internal energy. They primarily forced either rotation or compression component and reported cumulative energy transfers. Subsequent works by \citet{Miura:POF1995}, \citet{Graham:ApJ2010}, and \citet{Schmidt:PRE2019} advanced the field by investigating multiscale energy spectral transfers in compressive turbulence. In the following, we contrast our findings with these  studies, highlighting similarities and differences.

\citet{Graham:ApJ2010} investigated compressible magnetohydrodynamic turbulence using  $E_u(\mathbf{k}) =  \mathrm{Re}[\mathbf{v}(\mathbf{k}) \cdot \mathbf{u}^*(\mathbf{k})]/2$ as the modal kinetic energy and derived a dynamical equation for the energy transfers, which is Eq.~(\ref{eq:modal_ke_1}) of Sec.~\ref{sec:comp_eqns} of our paper. \citet{Graham:ApJ2010}  further decomposed the transfer term $T_u({\bf k})$ of Eq.~(\ref{eq:Tu}) as  
\be
T_u(\mathbf{k}) = T_{ua}(\mathbf{k}) + T_{uc}(\mathbf{k}),
\ee
where  
\bea
T_{ua}(\mathbf{k}) & = &-\frac{1}{2} \mathrm{Re}\left[ \mathbf{u}(\mathbf{k}) \cdot \widehat{[\mathbf{u} \cdot \nabla \mathbf{v}]}^*(\mathbf{k}) + \mathbf{v}^*(\mathbf{k}) \cdot \widehat{[\mathbf{u} \cdot \nabla \mathbf{u}]}(\mathbf{k}) \right], \\
T_{uc}(\mathbf{k}) & = & -\frac{1}{2} \mathrm{Re}\left[ \mathbf{u}(\mathbf{k}) \cdot \widehat{[\mathbf{v} \nabla \cdot \mathbf{u}]^*}(\mathbf{k}) \right],
\eea
with the hat symbol (e.g., $\widehat{\mathbf{u}}$) denoting the Fourier transform. They interpret $T_{ua}(\mathbf{k})$ as energy transfer due to advection, and $T_{uc}(\mathbf{k})$ as the transfer arising from compressibility effects; the latter term vanishes in the incompressible limit. These terms can be rewritten as  
\bea
T_{ua}(\mathbf{k}) & = & \frac{1}{2} \sum_{\mathbf{p}} \mathrm{Im} \big[ \{\mathbf{p} \cdot \mathbf{u}(\mathbf{q})\} \{\mathbf{v}(\mathbf{p}) \cdot \mathbf{u}^*(\mathbf{k}) + \mathbf{u}(\mathbf{p}) \cdot \mathbf{v}^*(\mathbf{k})\} \big], \\
T_{uc}(\mathbf{k}) & = & \frac{1}{2} \sum_{\mathbf{p}} \mathrm{Im} \big[ \{\mathbf{q} \cdot \mathbf{u}(\mathbf{q})\} \{\mathbf{v}(\mathbf{p}) \cdot \mathbf{u}^*(\mathbf{k})\} \big].
\eea
\citet{Schmidt:PRE2019} followed a similar approach to \citet{Graham:ApJ2010}, but used the density-weighted velocity $\mathbf{w} = \sqrt{\rho} \mathbf{u}$ to define the kinetic energy and its shell-to-shell spectral transfer function $\mathcal{T}_{uu}$. They derived that  
\bea
\mathcal{T}_{ua}({\bf k}) & = & \sum_{{\bf k} \in k} \sum_{ {\bf p} \in p} \mathrm{Im} \left[ \{\mathbf{p} \cdot \mathbf{u}(\mathbf{q})\} \{\mathbf{w}(\mathbf{p}) \cdot \mathbf{w}^*(\mathbf{k})\} \right], \\
\mathcal{T}_{uc}({\bf k}) & = & \frac{1}{2} \sum_{{\bf k} \in k} \sum_{ {\bf p} \in p} \mathrm{Im} \left[ \{\mathbf{q} \cdot \mathbf{u}(\mathbf{q})\} \{\mathbf{w}(\mathbf{p}) \cdot \mathbf{w}^*(\mathbf{k})\} \right].
\eea
The sum of ${T}_{ua}({\bf k})$ and ${T}_{uc}({\bf k})$, or $\mathcal{T}_{ua}({\bf k})$ and $\mathcal{T}_{uc}({\bf k})$, yields the kinetic energy flux. But, the advective and compressive terms of \citet{Graham:ApJ2010} and \citet{Schmidt:PRE2019}
fail to clearly distinguish energy transfers between rotational and compressive components. In contrast, our formalism achieves this distinction.

In the present paper, we go beyond Eq.~(\ref{eq:modal_ke_1}) and derive mode-to-mode energy transfer, $S^{uu}(\mathbf{k}|\mathbf{p}|\mathbf{q})$, which is kinetic energy transfer from wavenumber \textbf{p} to wavenumber \textbf{k} with mediation of wavenumber \textbf{q}. In addition, we separate the velocity field into rotational and compressional components, i.e., ${\bf u(k) = u}_R({\bf k}) + {\bf u}_C({\bf k})$, after which we derived mode-to-mode energy transfers for these components.
These transfers enable us to clearly derive energy transfers $T_{RR}$ (from $R$ to $R$), $T_{CC}$ (\textit{C} to \textit{C}), and $T_{CR}$ (\textit{R} to \textit{C}), and then energy fluxes among the rotational and compressive velocity components and the internal energy (See Figures~\ref{fig:mode_u} and \ref{fig:flux_comp}). Thus, our framework goes beyond the works of \citet{Graham:ApJ2010} and \citet{Schmidt:PRE2019} in a significant way.

To contrast our transfers terms with past works, we expand \citet{Graham:ApJ2010}'s $T_{ua}$ and $T_{uc}$ using the rotational and compressive components as
\bea
T_{ua}(\mathbf{k}) & = & \frac{1}{2} \sum_{\mathbf{p}} \mathrm{Im} \big[ \{\mathbf{p} \cdot \mathbf{u}(\mathbf{q})\} \{\mathbf{v}_{R}(\mathbf{p}) \cdot \mathbf{u}_{R}^*(\mathbf{k}) + \mathbf{v}_{R}(\mathbf{p}) \cdot \mathbf{u}_{C}^*(\mathbf{k}) + \mathbf{v}_{C}(\mathbf{p}) \cdot \mathbf{u}_{R}^*(\mathbf{k}) + \mathbf{v}_{C}(\mathbf{p}) \cdot \mathbf{u}_{C}^*(\mathbf{k})\} \big] \nonumber \\
&& + \frac{1}{2} \sum_{\mathbf{p}} \mathrm{Im} \big[ \{\mathbf{p} \cdot \mathbf{u}(\mathbf{q})\} \{\mathbf{u}_{R}(\mathbf{p}) \cdot \mathbf{v}_{R}^*(\mathbf{k}) + \mathbf{u}_{R}(\mathbf{p}) \cdot \mathbf{v}_{C}^*(\mathbf{k}) + \mathbf{u}_{C}(\mathbf{p}) \cdot \mathbf{v}_{R}^*(\mathbf{k}) + \mathbf{u}_{C}(\mathbf{p}) \cdot \mathbf{v}_{C}^*(\mathbf{k})\} \big], \\
T_{uc}(\mathbf{k}) & = & \frac{1}{2} \sum_{\mathbf{p}} \mathrm{Im} \big[ \{\mathbf{q} \cdot \mathbf{u}(\mathbf{q})\} \{\mathbf{v}_{R}(\mathbf{p}) \cdot \mathbf{u}_{R}^*(\mathbf{k}) + \mathbf{v}_{R}(\mathbf{p}) \cdot \mathbf{u}_{C}^*(\mathbf{k}) + \mathbf{v}_{C}(\mathbf{p}) \cdot \mathbf{u}_{R}^*(\mathbf{k}) + \mathbf{v}_{C}(\mathbf{p}) \cdot \mathbf{u}_{C}^*(\mathbf{k})\} \big],
\eea
Therefore, both $T_{ua}$ and $T_{uc}$ involve $T_{RR}$, $T_{CC}$, $T_{RC}$, and $T_{CR}$. Hence, they do not separate the energy transfers between rotational and compressive components. We overcome this difficulty in our framework using mode-to-mode energy transfers.  Note, however, that our formulas for pressure dilatation and viscous dissipation are similar to those of \citet{Graham:ApJ2010} and \citet{Schmidt:PRE2019}.


\citet{Aluie:PhysD2013} introduced a coarse-graining framework to analyze the kinetic energy cascade, which was later extended by  \citet{Wang:JFM2018} to separately examine the cascades of rotational and compressible kinetic energy. \citet{Banerjee:JFM2014} derived an exact relation for the two-point correlation function for compressible polytropic turbulence under the assumptions of statistical homogeneity. While these approaches have provided important insights into the nature of kinetic energy cascades in compressible turbulence, they do not fully resolve the detailed scale-to-scale energy transfers. 

\citet{Jagannathan:JFM2016} examined the scaling of global quantities with Mach and Reynolds numbers using numerical simulations of compressible turbulence with purely rotational forcing. In contrast, our simulations (see companion Letter~\cite{Singh:PRL2025_submitted}) employ mixed-mode forcing that excites both rotational and compressible components, resulting in different scaling behaviour. For example, while they observed that the compressive-to-rotational dissipation ratio $\epsilon_C / \epsilon_R \sim M_t^2$, our results indicate a weaker scaling of $\epsilon_C / \epsilon_R \approx \Pi_C / \Pi_R \sim M_t^{0.3}$~\cite{Singh:PRL2025_submitted}. This discrepancy may arise because of the differences in forcing, as \citet{Kida:JSC1990} and  \citet{Schmidt:PRE2019} reported significant variations in energy transfers and dissipation rates at similar Mach numbers but with different forcing schemes. \citet{Kida:JSC1990} and \citet{Jagannathan:JFM2016} showed that the properties of compressible turbulence vary with the nature of forcing.  We  need to explore how the energy fluxes vary
under different types of forcing and with the Mach numbers.
 
\citet{Kritsuk:Apj2007} showed that at high Mach numbers ($M_t \sim 6$), the velocity spectrum deviates from Kolmogorov scaling and approaches a Burgers-like behaviour. They proposed that the density-weighted velocity $\mathbf{v} = \rho^{1/3} \mathbf{u}$ restores Kolmogorov-like scaling in both velocity spectra and structure functions. Interestingly, they reported $k^{-3/2}$ scaling for both the rotational and compressive kinetic energy spectra. In contrast, our simulations~\cite{Singh:PRL2025_submitted} at lower Mach numbers ($M_t \in [0.15, 0.45]$) exhibit $k^{-5/3}$ scaling for rotational kinetic energy and $k^{-2}$ for the compressive component. These differences raise important questions about the evolution of energy cascades with increasing compressibility. To address this, we plan to extend our simulations to higher Mach numbers and analyze the corresponding energy fluxes, with the goal of bridging the gap between the subsonic and supersonic turbulence regimes and testing the persistence of these scaling trends.




The above discussion shows that our paper advances the energy transfers in compressible turbulence in a significant way.

\section{CONCLUSIONS}\label{sec:summary}

Despite extensive past research, a comprehensive understanding of compressible turbulence remains elusive, and a consistent description of energy fluxes has been lacking. This paper presents a significant advancement by introducing a novel framework for computing energy transfers and fluxes in compressible flows. Specifically, we derive mode-to-mode energy transfer rates in Fourier space and demonstrate detailed energy conservation within triads of interacting modes for the nonlinear terms. The formalism captures both kinetic and internal energy exchanges, thereby extending the spirit of incompressible turbulence analysis to the compressible regime.

We further decompose the kinetic energy into rotational and compressive components and derive their respective mode-to-mode transfer terms, including the cross-interaction contributions. Leveraging these results, we construct analytical expressions for rotational and compressive kinetic energy fluxes, cross-transfer fluxes between them, and transfers from kinetic to internal energy via pressure and viscous dissipation. This enables a detailed quantification of energy exchanges between rotational, compressive, and internal energy components. Notably, we also derive exact flux relations analogous to those in incompressible magnetohydrodynamic flows, demonstrating the universality of the mode-to-mode transfer formalism. This represents the first detailed and fully conservative analytical framework for energy transfers in compressible turbulence.

In the companion Letter~\cite{Singh:PRL2025_submitted}, we apply this framework to analyze data from direct numerical simulations of compressible turbulence at Mach numbers 0.15, 0.30, and 0.45, relevant to terrestrial and solar conditions. From the simulation data, we compute energy spectra, fluxes, and transfer rates across scales, and observe distinct scaling behaviours for the rotational and compressive components. These results underscore the utility of the formalism in analyzing multiscale energy dynamics in compressible flows.

Our work advances the study of compressible turbulence by providing a precise framework for analyzing interscale energy transfers. This opens avenues for exploring more complex flows, such as compressible convection and magnetohydrodynamics, with applications in planetary, stellar, and galactic environments. However, the current study, limited by isothermal conditions and a maximum Mach number of 0.45, constrains the results. Future work will expand these findings to higher Mach numbers and more general flow conditions, offering deeper insights into compressible turbulence.

\begin{acknowledgments}
The authors thank Rajesh Ranjan, Fahad Anwar, Sanjiva Lele, Parviz Moin, Manthan Verma, Abhishek Jha, Shashwat Nirgudkar, and Abhay for useful discussions. Simulations were performed on the HPC cluster of Kotak School of Sustainability (KSS), IITK.   LS thanks IITK for the Institute Postdoctoral Fellowship.  Part of this work was done in the Center for Turbulence Research, Stanford University, where MKV was a Visiting Senior Fellow. Part of this work was supported by Anusandhan National Research Foundation, India
(Grant Nos.~SERB/PHY/2021522 and SERB/PHY/2021473), and the J.~C.~Bose Fellowship (SERB/PHY/2023488).
\end{acknowledgments}

\appendix
\section{Derivation of dynamical equation for $E_u({\bf k})$ and combined energy transfers}
\label{app:ke_derv}
The momentum and velocity equations for a compressible flow, expressed in non-dimensional tensorial form, are given by
\be
\frac{\partial}{\partial t}(\rho u_i) + \frac{\partial}{\partial x_j} \left( \rho u_i u_j  + \delta_{ij} \sigma - \tau_{ij} \right) = \rho F_i,
\label{eq:momentum_app}
\ee
\be
\frac{\partial u_i}{\partial t} + u_j \frac{\partial u_i}{\partial x_j} + \frac{1}{\rho} \frac{\partial}{\partial x_j} \left( \delta_{ij} \sigma - \tau_{ij} \right) = F_i.
\label{eq:velocity}
\ee
Using a new field $\mathbf{v} = \rho \mathbf{u}$ and an application of Fourier transform to Eqs.~(\ref{eq:momentum_app}) and (\ref{eq:velocity}) yield
\bea
\frac{d}{dt} \mathbf{v}(\mathbf{k}) &=& -i \sum_{\mathbf{p}} \left\{ \mathbf{k} \cdot \mathbf{u}(\mathbf{q}) \right\} \mathbf{v}(\mathbf{p}) - i \mathbf{k} \sigma(\mathbf{k}) - \mathbf{d}(\mathbf{k}) + \mathbf{F'}(\mathbf{k}),
\label{eq:vk_ap}\\
\frac{d}{dt} \mathbf{u}(\mathbf{k}) &=& -i \sum_{\mathbf{p}} \left\{ \mathbf{p} \cdot \mathbf{u}(\mathbf{q}) \right\} \mathbf{u}(\mathbf{p}) - \tilde{\mathbf{\sigma}}(\mathbf{k}) - \tilde{\mathbf{d}}(\mathbf{k}) + \mathbf{F}(\mathbf{k}),
\label{eq:uk_ap}
\eea
where $\mathbf{k} = \mathbf{p} + \mathbf{q}$ and
\[
\tilde{\mathbf{\sigma}} = \frac{\nabla \sigma}{\rho}, \quad 
\mathbf{d} = - \partial_j \tau_{ij}, \quad 
\tilde{\mathbf{d}} = \frac{\mathbf{d}}{\rho}, \quad 
\mathbf{F'} = \rho \mathbf{F}.
\]
Taking the dot product of Eq.~(\ref{eq:vk_ap}) with $\mathbf{u}^{*}(\mathbf{k})$ and Eq.~(\ref{eq:uk_ap}) with $\mathbf{v}^{*}(\mathbf{k})$ yields:
\bea
\mathbf{u}^{*}(\mathbf{k}) \cdot \frac{d}{dt} \mathbf{v}(\mathbf{k}) &=& 
-i \sum_{\mathbf{p}} \left\{ \mathbf{k} \cdot \mathbf{u}(\mathbf{q}) \right\} \left\{ \mathbf{v}(\mathbf{p}) \cdot \mathbf{u}^{*}(\mathbf{k}) \right\}
- i \sigma(\mathbf{k}) \left\{ \mathbf{k} \cdot \mathbf{u}^{*}(\mathbf{k}) \right\}
- \mathbf{d}(\mathbf{k}) \cdot \mathbf{u}^{*}(\mathbf{k})
+ \mathbf{F'}(\mathbf{k}) \cdot \mathbf{u}^{*}(\mathbf{k}),
\label{eq:u*vk}\\
\mathbf{v}^{*}(\mathbf{k}) \cdot \frac{d}{dt} \mathbf{u}(\mathbf{k}) &=& 
-i \sum_{\mathbf{p}} \left\{ \mathbf{p} \cdot \mathbf{u}(\mathbf{q}) \right\} \left\{ \mathbf{u}(\mathbf{p}) \cdot \mathbf{v}^{*}(\mathbf{k}) \right\}
- \tilde{\mathbf{\sigma}}(\mathbf{k}) \cdot \mathbf{v}^{*}(\mathbf{k})
- \tilde{\mathbf{d}}(\mathbf{k}) \cdot \mathbf{v}^{*}(\mathbf{k})
+ \mathbf{F}(\mathbf{k}) \cdot \mathbf{v}^{*}(\mathbf{k}).
\label{eq:v*uk}
\eea
By adding Eqs.~(\ref{eq:u*vk}) and (\ref{eq:v*uk}) we obtain
    \bea
    \mathbf{u}^{*}(\mathbf{k}) \cdot \frac{d}{dt} \mathbf{v}(\mathbf{k}) + \mathbf{v}^{*}(\mathbf{k}) \cdot \frac{d}{dt} \mathbf{u}(\mathbf{k})
    & = & -i \sum_{\mathbf{p}} \left\{ \mathbf{k} \cdot \mathbf{u}(\mathbf{q}) \right\} \left\{ \mathbf{v}(\mathbf{p}) \cdot \mathbf{u}^{*}(\mathbf{k}) \right\}  -i \sum_{\mathbf{p}} \left\{ \mathbf{p} \cdot \mathbf{u}(\mathbf{q}) \right\} \left\{ \mathbf{u}(\mathbf{p}) \cdot \mathbf{v}^{*}(\mathbf{k}) \right\} \nonumber \\
&& - \big[i \sigma(\mathbf{k}) \left\{ \mathbf{k} \cdot \mathbf{u}^{*}(\mathbf{k}) \right\} + \tilde{\mathbf{\sigma}}(\mathbf{k}) \cdot \mathbf{v}^{*}(\mathbf{k})\big]  - \big[\mathbf{d}(\mathbf{k}) \cdot \mathbf{u}^{*}(\mathbf{k}) + \tilde{\mathbf{d}}(\mathbf{k}) \cdot \mathbf{v}^{*}(\mathbf{k})\big] \nonumber \\
&& + \big[\mathbf{F'}(\mathbf{k}) \cdot \mathbf{u}^{*}(\mathbf{k}) + \mathbf{F}(\mathbf{k}) \cdot \mathbf{v}^{*}(\mathbf{k})\big].
\label{eq:vu}
    \eea
We further add Eq.~(\ref{eq:vu}) with its complex conjugate and obtain [see Eq.~(\ref{eq:modal_ke_1})]:
\be
\frac{d}{dt} E_u(\mathbf{k}) = T_u(\mathbf{k}) - Q_{I,u}(\mathbf{k}) - D_{I,u}(\mathbf{k}) + \mathcal{F}_u(\mathbf{k}).
\label{eq:modal_ke_1_app}
\ee
The terms on the right-hand side represent the nonlinear transfer, pressure dilatation, viscous dissipation, and energy injection by ${\bf F}$, respectively; these terms are defined in Eqs.~(\ref{eq:Tu}-\ref{eq:mathcal_F}) of the main text.

For a single triad $({\bf k',p,q})$ with ${\bf k'+p+q}=0$~\cite{Dar:PD2001,Verma:book:ET}, the energy equation is
\be
\frac{d}{dt} E_u(\mathbf{k'}) = S^{uu}(\mathbf{k'|p,q}) - Q_{I,u}(\mathbf{k'}) 
 - D_{I,u}(\mathbf{k'}) +  \mathcal{F}_u(\mathbf{k'}) 
\label{eq:modal_ke_2_app},
\ee
where
\bea
S^{uu}(\mathbf{k}'|\mathbf{p}, \mathbf{q}) = -\frac{1}{2} \mathrm{Im}\big[\{\mathbf{k}' \cdot \mathbf{u}(\mathbf{q})\} \{\mathbf{v}(\mathbf{p}) \cdot \mathbf{u}(\mathbf{k}')\} 
- \{\mathbf{p} \cdot \mathbf{u}(\mathbf{q})\} \{\mathbf{u}(\mathbf{p}) \cdot \mathbf{v}(\mathbf{k}')\}\big] + \mathbf{p} \leftrightarrow \mathbf{q}
\label{eq:combined_ET_k_app}
\eea
is the \textit{combined kinetic energy transfer} to wavenumber $\mathbf{k'}$ from wavenumbers \textbf{p} and  \textbf{q}. Similarly, we can write the combined kinetic energy transfers to wavenumbers $\mathbf{p}$ and $\mathbf{q}$ as
\begin{gather}
S^{uu}(\mathbf{p}|\mathbf{k}', \mathbf{q}) = -\frac{1}{2} \mathrm{Im}\big[\{\mathbf{p} \cdot \mathbf{u}(\mathbf{q})\} \{\mathbf{v}(\mathbf{k'}) \cdot \mathbf{u}(\mathbf{p})\} 
- \{\mathbf{k}' \cdot \mathbf{u}(\mathbf{q})\} \{\mathbf{u}(\mathbf{k}') \cdot \mathbf{v}(\mathbf{p})\}\big] + \mathbf{k}' \leftrightarrow \mathbf{q},
\label{eq:combined_ET_p_app} \\
S^{uu}(\mathbf{q}|\mathbf{p}, \mathbf{k}') = -\frac{1}{2} \mathrm{Im}\big[\{\mathbf{q} \cdot \mathbf{u}(\mathbf{k}')\} \{\mathbf{v}(\mathbf{p}) \cdot \mathbf{u}(\mathbf{q})\} 
- \{\mathbf{p} \cdot \mathbf{u}(\mathbf{k}')\} \{\mathbf{u}(\mathbf{p}) \cdot \mathbf{v}(\mathbf{q})\}\big] + \mathbf{p} \leftrightarrow \mathbf{k'}.
\label{eq:combined_ET_q_app}
\end{gather}
An addition of Eqs.~(\ref{eq:combined_ET_k_app},\ref{eq:combined_ET_p_app}, \ref{eq:combined_ET_q_app}) yields
    \bea
    S^{uu}({\bf k'|p, q}) + S^{uu}({\bf p|k', q}) + S^{uu}({\bf q|p, k})
    & = & -\frac{1}{2} \mathrm{Im}\big[\{\mathbf{k}' \cdot \mathbf{u}(\mathbf{q})\} \{\mathbf{v}(\mathbf{p}) \cdot \mathbf{u}(\mathbf{k}')\} 
- \{\mathbf{p} \cdot \mathbf{u}(\mathbf{q})\} \{\mathbf{u}(\mathbf{p}) \cdot \mathbf{v}(\mathbf{k}')\}\big]  \nonumber \\
&& -\frac{1}{2} \mathrm{Im}\big[\{\mathbf{k}' \cdot \mathbf{u}(\mathbf{p})\} \{\mathbf{v}(\mathbf{q}) \cdot \mathbf{u}(\mathbf{k}')\} 
- \{\mathbf{q} \cdot \mathbf{u}(\mathbf{p})\} \{\mathbf{u}(\mathbf{q}) \cdot \mathbf{v}(\mathbf{k}')\}\big]  \nonumber \\
&& -\frac{1}{2} \mathrm{Im}\big[\{\mathbf{p} \cdot \mathbf{u}(\mathbf{q})\} \{\mathbf{v}(\mathbf{k'}) \cdot \mathbf{u}(\mathbf{p})\} 
- \{\mathbf{k}' \cdot \mathbf{u}(\mathbf{q})\} \{\mathbf{u}(\mathbf{k}') \cdot \mathbf{v}(\mathbf{p})\}\big] \nonumber \\
&& -\frac{1}{2} \mathrm{Im}\big[\{\mathbf{p} \cdot \mathbf{u}(\mathbf{k}')\} \{\mathbf{v}(\mathbf{q}) \cdot \mathbf{u}(\mathbf{p})\} 
- \{\mathbf{q} \cdot \mathbf{u}(\mathbf{k}')\} \{\mathbf{u}(\mathbf{q}) \cdot \mathbf{v}(\mathbf{p})\}\big] \nonumber \\
&& -\frac{1}{2} \mathrm{Im}\big[\{\mathbf{q} \cdot \mathbf{u}(\mathbf{k}')\} \{\mathbf{v}(\mathbf{p}) \cdot \mathbf{u}(\mathbf{q})\} 
- \{\mathbf{p} \cdot \mathbf{u}(\mathbf{k}')\} \{\mathbf{u}(\mathbf{p}) \cdot \mathbf{v}(\mathbf{q})\}\big] \nonumber \\
&& -\frac{1}{2} \mathrm{Im}\big[\{\mathbf{q} \cdot \mathbf{u}(\mathbf{p})\} \{\mathbf{v}(\mathbf{k}') \cdot \mathbf{u}(\mathbf{q})\} 
- \{\mathbf{k}' \cdot \mathbf{u}(\mathbf{p})\} \{\mathbf{u}(\mathbf{k}') \cdot \mathbf{v}(\mathbf{q})\}\big]
\big] \nonumber \\
& = & 0, 
    \eea
which is the statement of \textit{detailed conservation law} for compressible hydrodynamics.

\section{Mode-to-mode energy transfers in terms of $\mathbf{w} = \sqrt{\rho} \mathbf{u}$}
\label{app:mode_w}

\citet{Kida:JSC1990, Kida:JSC1992},
\citet{Miura:POF1995},  \citet{Schmidt:PRE2019}, and \citet{Grete:PP2017} employed  the density-weighted velocity field $\mathbf{w} = \sqrt{\rho} \mathbf{u}$ for deriving energy spectra and transfers in compressible turbulence. In terms of $\mathbf{w}$, the kinetic energy density 
\be
E_u = \frac{1}{2} |\mathbf{w}|^2
\ee
is quadratic that helps in deriving spectral energy density.
 The dynamical equation for ${\bf w}$ is
\be
\frac{d}{dt} {\bf w}
= \frac{\partial {\bf w}}{\partial t}  + ({\bf u} \cdot \nabla){\bf w}
= \sqrt{\rho}\frac{\partial {\bf u}}{\partial t} + \frac{{\bf u}}{2\sqrt{\rho}}\frac{\partial \rho}{\partial t} + \sqrt{\rho}({\bf u} \cdot \nabla){\bf u} + \frac{{\bf u}}{2\sqrt{\rho}} ({\bf u} \cdot \nabla)\rho.
\label{eq:momentum_w_mid}
\ee
Using Eqs.~(\ref{eq:continuity},\ref{eq:velocity}), we can rewrite Eq.~(\ref{eq:momentum_w_mid}) as
\be
\frac{\partial {\bf w}}{\partial t}
= - ({\bf u} \cdot \nabla){\bf w}
- \frac{1}{2} {\bf w} (\nabla \cdot {\bf u})
- \frac{1}{\sqrt{\rho}} \nabla \sigma 
+ \frac{1}{\mathrm{Re}_0\sqrt{\rho}} [\nabla^2 {\bf u} + \frac{1}{3}\nabla(\nabla \cdot {\bf u})] + \sqrt{\rho} {\bf F},
\label{eq:momentum_w}
\ee
whose Fourier transformation is
\be
\frac{d}{dt} \mathbf{w}(\mathbf{k}) =
- i \sum_{\mathbf{p}} \{\mathbf{p} \cdot \mathbf{u}(\mathbf{q})\} \mathbf{w}(\mathbf{p})
- \frac{i}{2} \sum_{\bf{p}} \{\mathbf{q} \cdot \mathbf{u}(\mathbf{q})\} \mathbf{w}(\mathbf{p}) - \overline{\pmb{\sigma}}(\mathbf{k}) - \overline{\mathbf{d}}(\mathbf{k}) + \overline{\mathbf{F}}(\mathbf{k}), \label{eq:fourier}
\ee
where $\mathbf{k} = \mathbf{p} + \mathbf{q}$ and
\be
\overline{\pmb{\sigma}} = \frac{\nabla \sigma}{\sqrt{\rho}}, ~~~
\overline{\mathbf{d}} = -\frac{1}{\mathrm{Re}_0\sqrt{\rho}} [\nabla^2 {\bf u} + \frac{1}{3}\nabla(\nabla \cdot {\bf u})],~~~
\overline{\mathbf{F}} = \sqrt{\rho} \mathbf{F}.
\ee
Equation~(\ref{eq:fourier}) can be rearranged as
\be
\frac{d}{dt} \mathbf{w}(\mathbf{k}) =
- \frac{i}{2} \sum_{\mathbf{p}} \{(\mathbf{k} + \mathbf{p}) \cdot \mathbf{u}(\mathbf{q})\} \mathbf{w}(\mathbf{p}) - \overline{\pmb{\sigma}}(\mathbf{k}) - \overline{\mathbf{d}}(\mathbf{k}) + \overline{\mathbf{F}}(\mathbf{k}) .\label{eq:fourier_w}
\ee
We take the dot product of Eq.~(\ref{eq:fourier_w}) with $\mathbf{w}^*(\mathbf{k})$ that yields
\be
\mathbf{w}^*(\mathbf{k}) \cdot \frac{d}{dt} \mathbf{w}(\mathbf{k}) =
- \frac{i}{2} \sum_{\mathbf{p}} \{(\mathbf{k} +  \mathbf{p})\cdot \mathbf{u}(\mathbf{q})\} \{\mathbf{w}(\mathbf{p}) \cdot \mathbf{w}^*(\mathbf{k})\} - [\overline{\pmb{\sigma}}(\mathbf{k}) + \overline{\mathbf{d}}(\mathbf{k}) - \overline{\mathbf{F}}(\mathbf{k})] \cdot \mathbf{w}^*(\mathbf{k}). \label{eq:w_wstar}
\ee
After adding the Eq.~(\ref{eq:w_wstar}) with its complex conjugate, we obtain the dynamical equation for the modal kinetic energy $|{\bf w(k)}|^2/2$ as 
\be
\frac{d}{d t} E_u (\mathbf{k}) = \overline{T}_u (\mathbf{k}) - \overline{Q}_{I,u} (\mathbf{k}) - \overline{D}_{I,u}(\mathbf{k}) + \overline{\mathcal{F}}_u (\mathbf{k}).\label{eq:modal_ke_w}
\ee
Here,
\bea 
\overline{T}_u(\mathbf{k}) & = & \frac{1}{2} \sum_{\mathbf{p}} \mathrm{Im} \big[ \{(\mathbf{k} + \mathbf{p})\cdot \mathbf{u}(\mathbf{q})\} \{\mathbf{w}(\mathbf{p}) \cdot \mathbf{w}^*(\mathbf{k})\} \big], \label{eq:Tu_w} \\
\overline{Q}_{I,u}(\mathbf{k}) & = &
 \mathrm{Re} \big[\overline{\pmb{\sigma}}(\mathbf{k}) \cdot \mathbf{w}^*(\mathbf{k})\big], \label{eq:Qu_w} \\
\overline{D}_{I,u}(\mathbf{k}) & = & \mathrm{Re} \big[ {\overline{\mathbf{d}}}(\mathbf{k}) \cdot \mathbf{w}^*(\mathbf{k}) \big],\label{eq:Du_w} \\
\overline{\mathcal{F}}_u(\mathbf{k}) & = & \mathrm{Re}\big[ {\overline{\mathbf{F}}}(\mathbf{k}) \cdot \mathbf{w}^*(\mathbf{k}) \big]. \label{eq:mathcal_F_w}
\eea

For a single triad $({\bf k',p,q})$ with ${\bf k'+p+q}=0$ (${\bf k'= -k}$)~\cite{Dar:PD2001,Verma:book:ET}, 
\be
\frac{d}{d t} E_u(\mathbf{k'}) = \overline{S}^{uu}(\mathbf{k'|p,q}) - \overline{Q}_{I,u}(\mathbf{k'}) 
 - \overline{D}_{I,u}(\mathbf{k'}) +  \overline{\mathcal{F}}_u(\mathbf{k'}),
\label{eq:modal_ke_w1}
\ee
where
\bea
\overline{S}^{uu}(\mathbf{k}'|\mathbf{p}, \mathbf{q}) = -\frac{1}{2} \mathrm{Im}\big[\{(\mathbf{k}' -  \mathbf{p})\cdot \mathbf{u}(\mathbf{q})\} \{\mathbf{w}(\mathbf{p}) \cdot \mathbf{w}(\mathbf{k}')\}\big] + \mathbf{p} \leftrightarrow \mathbf{q}
\label{eq:combined_ET_w}
\eea
is the \textit{combined KE transfer} to wavenumber $\mathbf{k'}$ from wavenumbers \textbf{p} and  \textbf{q}. Following the same steps as in Sec.~\ref{sec:mode_to_mode}, we derive 
the mode-to-mode energy transfers for this framework. Here, 
\be
\overline{S}^{uu}(\mathbf{k'}|\mathbf{p}|\mathbf{q}) = -\frac{1}{2} \mathrm{Im}\big[\{(\mathbf{k'} - \mathbf{p})\cdot \mathbf{u}(\mathbf{q})\} \{\mathbf{w}(\mathbf{p}) \cdot \mathbf{w}(\mathbf{k'})\}\big].
\label{Eq:M2M_uu_w}
\ee
is the \textit{\mm} KE transfer from wavenumber \textbf{p} to wavenumber $\bf{k'}$ with the mediation  of wavenumber \textbf{q}. 
It can be shown that
\bea
\overline{S}^{uu}({\bf k'|p|q}) + \overline{S}^{uu}({\bf k'|q|p}) & = & \overline{S}^{uu}({\bf k'|p,q}),\label{Eq:M2M_P1_w}\\
\overline{S}^{uu}({\bf k'|p|q}) + \overline{S}^{uu}({\bf p|k'|q}) & = & 0, \label{Eq:M2M_P2_w}
\eea 
which have the same form as in Eqs.~(\ref{Eq:M2M_P1},\ref{Eq:M2M_P2}). 

Thus, we derive mode-to-mode KE transfers for compressible turbulence with $\mathbf{w} = \sqrt{\rho}\mathbf{u}$ variable. The formulas for the energy fluxes in terms of $\overline{S}^{uu}({\bf k'|p|q})$ are same as those with $S^{uu}({\bf k'|p|q})$, except that $S \rightarrow \overline{S}$. We verified that the numerical energy fluxes using the new formulas are same as those described in Sec.~\ref{sec:Fluxes}.

\bibliography{bib/journal, bib/book, bib/compress_journal, bib/compress_book, bib/thesis, bib/conf, bib/conf_long}

\end{document}